\documentclass[twocolumn,showpacs,aps,amssymb,floatfix,prd,amsmath,preprintnumbers]{revtex4}
\usepackage{epstopdf}
\usepackage{capt-of}
\usepackage{graphicx}  % Include figure files
\usepackage{dcolumn}   % Align table columns on decimal point
\usepackage{float}
\usepackage{hyperref}
\begin{document}
%\input epsf.tex
%%%%%%%%%%%%
%%%%%%%%%%%
\title{An exponential shape function for wormholes in modified gravity}

\author{ P.H.R.S. Moraes$^{1,2}$\footnote{Email:  moraes.phrs@gmail.com},
  P.K. Sahoo$^{3}$\footnote{Email: pksahoo@hyderabad.bits-pilani.ac.in},
  Shreyas Sunil Kulkarni $^{3}$\footnote{Email: sskshreyas@gmail.com},
Shivaank Agarwal$^{3}$\footnote{Email: shivaank.agarwal@gmail.com}}\
  
\affiliation{$^{1}$Universit\`a degli studi di Napoli ``Federico II'' - Dipartimento di Fisica, Napoli I-80126, Italy}
\affiliation{$^{2}$ ITA - Instituto Tecnol\'ogico de Aeron\'autica - Departamento de F\'isica, 12228-900, S\~ao Jos\'e dos Campos, S\~ao Paulo, Brasil}
\affiliation{$^{3}$ Department of Mathematics, Birla Institute of
Technology and Science-Pilani, Hyderabad Campus, Hyderabad-500078,
India}

%%%%%%%%%%%%
\begin{abstract}

Here, we propose a new exponential shape function in wormhole geometry within modified gravity. The energy conditions and equation of state parameter are obtained. The radial and tangential null energy conditions are validated, as well as the weak energy condition, which indicates the absence of exotic matter due to modified gravity allied with such a new proposal.

\end{abstract}



\maketitle

%%%%%%%%%%%%%%%%%%%%%%%%%%%%%%%%%%%%%%%%%%

\section{Introduction}\label{I}

Wormholes (WHs) are asymptotically flat tube-like structures. They are said to be useful for interstellar travel as they could connect two different points in the same universe or two points in different universe \cite{morris/1988}. 

Wormholes arise from the solutions of General Relativity (GR). Schwarzschild's WH was the first WH-like solution to be obtained \cite{flamm/1916}. It was later discovered that it would collapse very quickly, preventing it to be traversable \cite{fuller/1962}.

This issue was delved deeper in \cite{morris/1988,yurtzever/1988}, in which a static and spherically symmetric metric was suggested to describe WHs and the required energy constraints to make them traversable were discussed. This analysis led to the violation of the null energy condition (NEC), so that in order to the GRWH to be traversable, it should be filled by exotic matter (matter violating the NEC). 

The issue is that finding suitable contenders for exotic matter has never been done. Therefore, modified gravity theories (MGTs), that input some extra degrees of freedom to GR in a fundamental level, appear as a possibility to treat this issue, by addressing the question of whether is possible to have stable WH solutions with no need for exotic matter. On the regard of MGTs, we recommend the following important reviews \cite{nojiri/2017}-\cite{capozziello/2011}.

Naturally, MGTs have been used to address not only the exotic matter issue, but other several issues of current observational astrophysics and cosmology \cite{jain/2013}-\cite{amarzguioui/2006}.

Due to the lack of WH observations so far, despite all the efforts and proposals \cite{rahaman/2014}-\cite{jusufi/2018}, some geometrical and material features of WHs, such as the shape function and equation of state (EoS), are still not precisely known. Particularly, several forms for the shape function $b(r)$ have been proposed and analysed so far, as one can check, for instance, \cite{konoplya/2018}-\cite{godani/2019}. In the present article we will propose a new form for the WH shape function. As $b(r)$ is not arbitrary and have to obey several conditions, which will be presented some sections below, our proposal must be, naturally, in accordance with these conditions. 

We will also aim the obedience of the WH energy conditions (ECs). In order to attain that, allied to the shape function proposed we will underline our model with a particular MGT named $f(R,T)$ theory \cite{harko/2011}. The $f(R,T)$ theory starts from a gravitational action that substitutes the Ricci scalar $R$ in the usual Einstein-Hilbert action by a general function of $R$ and $T$, with $T$ being the trace of the energy momentum-tensor $T_{ij}$. The motivation to insert some material terms in the gravitational action is related to the possible existence of imperfect fluids in the universe. Since WHs material content is described by an anisotropic fluid, their investigation in such a theory of gravity is well motivated. 

The $f(R,T)$ gravity authors have argued that due to the coupling of matter and geometry, this gravity model depends on a source term, which is nothing but the variation of the matter stress-energy tensor \cite{harko/2011}. This source term could be related to quantum effects since it could lead to a particle creation scenario \cite{harko/2014}. As a result, the motion of test particles in $f(R,T)$ gravity is not along geodesic path due to the presence of an extra force perpendicular to the four-velocity. 

\section{The $f(R,T)$ theory of gravity}\label{II}

In order to obtain our WH solutions, we will consider the $f(R,T)$ modified theory of gravity \cite{harko/2011}, where the gravitational Lagrangian is given by an arbitrary function of $R$ and $T$. The gravitational action for this theory is defined, then, as \cite{harko/2011}

\begin{equation}\label{e1}
S=\frac{1}{16\pi}\int d^{4}x\sqrt{-g}f(R,T)+\int d^{4}x\sqrt{-g}\mathcal{L}_m.
\end{equation}
In Eq.(\ref{e1}), $f(R,T)$ is an arbitrary function of $R$ and $T$, $g$ denotes the determinant of the metric $g_{ij}$ and $\mathcal{L}_m$ is the matter lagrangian. Moreover, we are assuming natural units.

By varying Eq.(\ref{e1}) with respect to the metric $g_{ij}$, the field equations obtained are
\begin{multline}\label{e4}
f_R(R,T)R_{ij}-\frac{1}{2}f(R,T)g_{ij}+(g_{ij}\square -\nabla_i\nabla_j)f_R(R,T)\\ =8\pi T_{ij}-f_T(R,T)T_{ij}-f_T(R,T)\Theta_{ij}.
\end{multline}
Here, $f_R(R,T)=\partial f(R,T)/\partial R$, $f_T(R,T)=\partial f(R,T)/\partial T$,
\begin{equation}
T_{\mu\nu}=-\frac{2}{\sqrt{-g}}\frac{\delta(\sqrt{-g})\mathcal{L}_m}{\delta g^{\mu\nu}}
\end{equation}
 and
\begin{equation}\label{e5}
\Theta_{ij}=-2T_{ij}-p g_{ij}
\end{equation}
if we choose $\mathcal{L}_m=-p$, where $p$ is the total pressure of the fluid. $\mathcal{L}_m=-p$ represents the matter lagrangian density of a perfect fluid, which, in fact, is not uniquely defined \cite{harko/2011,harko/2014b,avelino/2018,moraes/2019}. It is quite usual to see choices such as $\mathcal{L}_m=\rho$, with $\rho$ being the matter-energy density, and $\mathcal{L}_m=-p$, but even $\mathcal{L}_m=T$ was already used \cite{avelino/2018}. It is known that geometry-matter coupling gravity theories, such as the $f(R,T)$ gravity, predict the existence of an extra force acting orthogonally to the four-velocity in a (non-)geodesic motion. This extra force remarkably depends on the matter lagrangian density and vanishes if $\mathcal{L}_m=-p$ \cite{harko/2014b,bertolami/2008}, which is the reason why we have assumed so. 

We will consider $f(R,T)=R+2\lambda T$, which was assumed as the functional form for the function $f(R,T)$ in several approaches such as \cite{smsb/2018}-\cite{dkmkmr/2019}, among many others. The considered form is the simplest one, which reduces to general
relativity for the choice of $\lambda = 0$. For this choice one can easily correlate
the obtained results with the most successful Einstein's general relativity.The $f(R,T)$ gravity field equations, for this case, read
\begin{equation}\label{e7}
G_{ij}=8\pi T_{ij}+\lambda Tg_{ij}+2\lambda(T_{ij}+pg_{ij}),
\end{equation}
with $G_{ij}$ being the Einstein tensor.

\section{The wormhole field equations}\label{III}

In order to describe the geometry of WHs space-time we use the modified version of the spherically symmetric space-time metric as

\begin{equation}\label{e8}
ds^2 = -dt^2 + \left[\frac{dr^2}{1- \frac{b(r)}{r}}+ r^2 d\theta^2 +r^2\sin^2\theta d\phi^2\right] 
\end{equation}
Here, the redshift function was normalized. Constant redshift functions were assumed in several references, such as \cite{jamil/2010}-\cite{elizalde/2019}. The radial coordinate $r$ decreases from $\infty$ to a minimum value $r_0$, called the WH throat, and then increases to $\infty$. The shape function $b(r)$ needs to satisfy the following conditions:

$\bullet$ At the throat, $b(r_0)=r_0$ and for $r>r_0$, $1-\frac{b(r)}{r}>0$;

$\bullet$ $b'(r_0)<1$ (flaring-out condition);

$\bullet$ $\lim_{r \to \infty} \frac{b(r)}{r} = 0$ (asymptotically flatness condition);
with primes indicating radial derivatives.

The Ricci scalar for the WH metric is obtained as

\begin{equation}\label{e9}
R = \frac{2b'}{r^2}.
\end{equation}

The non-zero Einstein tensor components for the WH metric are

\begin{equation}\label{e10}
\begin{split}
G_{00} &=\frac{b'}{r^2}, \\
G_{11} &= -\frac{b}{r^2 (r-b)}, \\
G_{22} &= \frac{b-r b'}{2 r}, \\
G_{33} &= \frac{\sin ^2\theta \left(b-r b'\right)}{2 r}. \\
\end{split}
\end{equation}

The field equations (\ref{e7}) for the metric (\ref{e8}) and anisotropic energy-momentum tensor are, then, written explicitly as

\begin{equation}\label{e11}
\begin{split}
\frac{b'}{r^2} &= (8\pi + \lambda)\rho - \lambda(p_r + 2p_t),\\
- \frac{b}{r^3} &= (8\pi + 3\lambda)p_r + \lambda(\rho+2p_t),\\
 \frac{b-rb'}{2 r^3} &= (8\pi + 4\lambda)p_t+\lambda(\rho+p_r),
\end{split}
\end{equation}
with $p_r$ and $p_t$ being, respectively, the radial and transverse pressures of the WH, such that the WH total pressure is $p=(p_r+2p_t)/3$. One can obtain the values in GR by making $\lambda=0$ in the above equation.

\section{Energy Conditions}\label{IV}

In GR, the energy conditions are a set of inequalities that are required to prove various important theorems such as those related to WHs and black holes. It is well known that static traversable GRWHs violate the energy conditions near the WH throat \cite{morris/1988,Hochberg/1997}.

The ECs have significant theoretical applications, suchlike the Hawking Penrose singularity conjecture, which is based on the (strong energy condition) SEC \cite{Hawking/1973} while the (dominant energy condition) DEC is applicable to proof the positive mass theorem \cite{Schoen/1981}. Further, the NEC is a basic requirement to derive the second law of black hole thermodynamics \cite{Carroll/2004}. The cosmological terms suchlike deceleration, look back time, distance modulus and statefinder parameters are seen in terms of redshift using ECs in \cite{Visser/1997a}. 

The ECs have been studied in MGTs, like $f(R)$ gravity, Brans-Dicke theory, $f(G)$ gravity, $f(G,\mathcal{T})$ gravity \cite{Santos/2007}-\cite{Sharif/2016}, with $\mathcal{T}$ being the torsion scalar. The generalized ECs are analyzed in MGTs considering the degrees of freedom related to scalar fields and curvature invariants \cite{capozziello/2014,capozziello/2015}. In particular, the ECs were derived for a power law solution in $f(R,T)$ gravity and the stability of the same were established \cite{Sharif/2013}.

We will take into consideration the following energy conditions for a perfect fluid \cite{visser/1995}

\begin{enumerate}
\item  Weak energy condition (WEC):
$$\rho \geq 0; \rho + p_i > 0; $$
\item  NEC:
$$\rho + p_i \geq 0; $$
\item  SEC:
$$\rho + p_r + 2p_t \geq 0; $$
\item  DEC:
$$ \rho \geq |p_i|,$$
\end{enumerate}
where $i=r,t$.

In the following we will construct the above energy conditions for the $f(R,T)$ WHs presented in Section \ref{III} for a new proposal for the shape function, namely an exponential shape function. 

\subsection{Exponential shape function}\label{VI}

In this section we propose a new exponential form for the shape function as

\begin{equation}\label{e14}
b(r)= r_0 \cdot  e^{1-\frac{r}{r_0}}.
\end{equation}

\begin{figure}[H]
\includegraphics[scale =0.28]{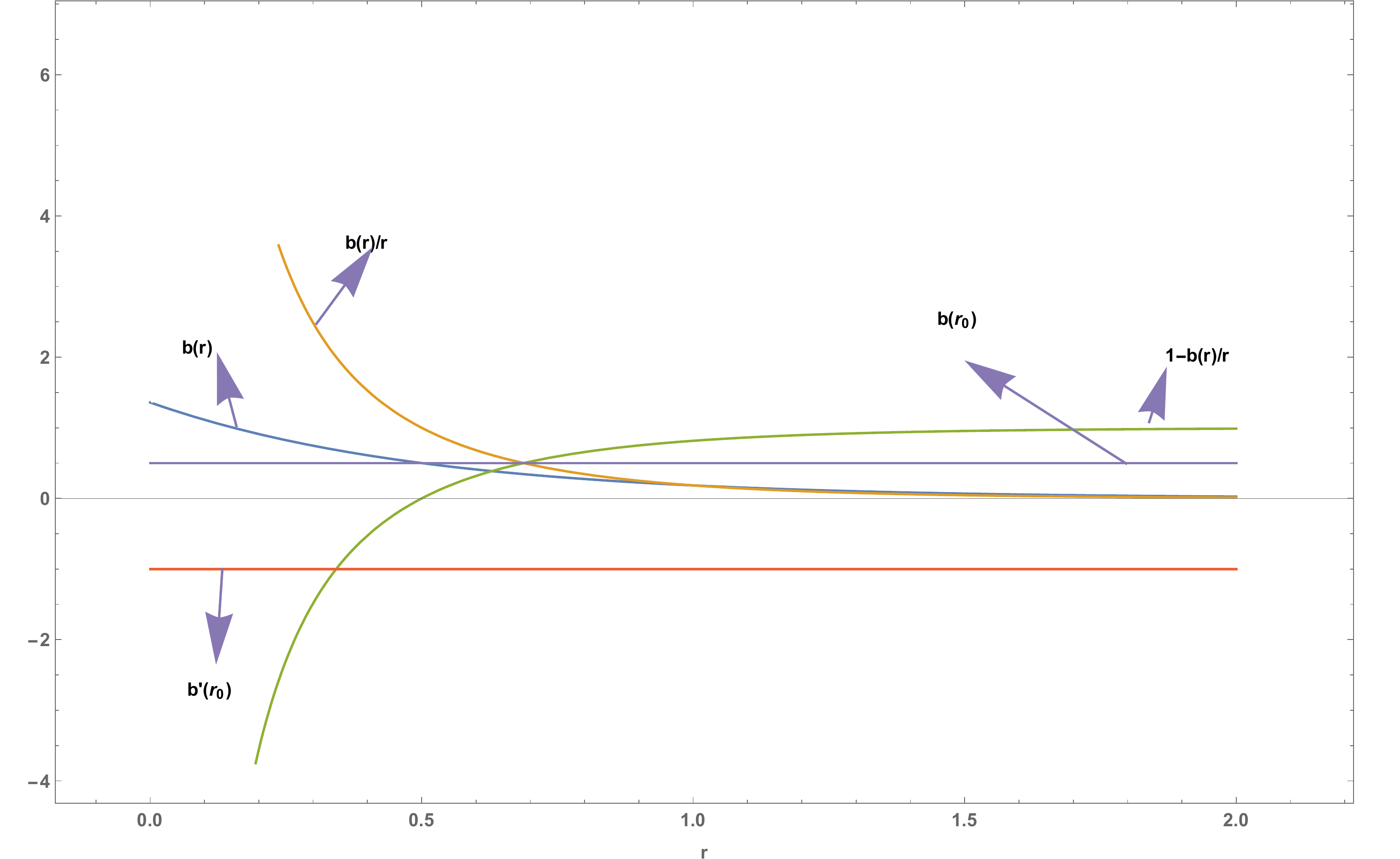}
\caption{Features of the shape function \eqref{e14} with $r_0=0.5$.}\label{fig9}
\end{figure}

In Fig.\ref{fig9} below we can see some features of $b(r)$ \eqref{e14}. 

One can observe from Fig.\ref{fig9} that the shape function satisfies all the basic requirements given in Section \ref{III}. %Hence, the proposed shape function can be used to derive the solutions for the matter-energy density and pressures.

Using the above shape function in the field equations (\ref{e11}), we obtain 

\begin{equation}\label{e15}
\begin{split}
\rho &= - \frac{e^{1-\frac{r}{r_0}}}{2 (\lambda +4 \pi ) r^2},\\
p_r &=- \frac{e^{1-\frac{r}{{r_0}}} [4 (\pi -1) r+(\lambda +4 \pi ) r_0]}{2 (\lambda +4 \pi )^2 r^3},\\
p_t &=\frac{e^{1-\frac{r}{r_0}} (r+r_0)}{4 (\lambda +4 \pi ) r^3}.\\
\end{split}
\end{equation}

Below we plot the radial EoS as well as the energy conditions for the present WH model with exponential shape function.

\begin{figure}[H]
\includegraphics[scale =0.35]{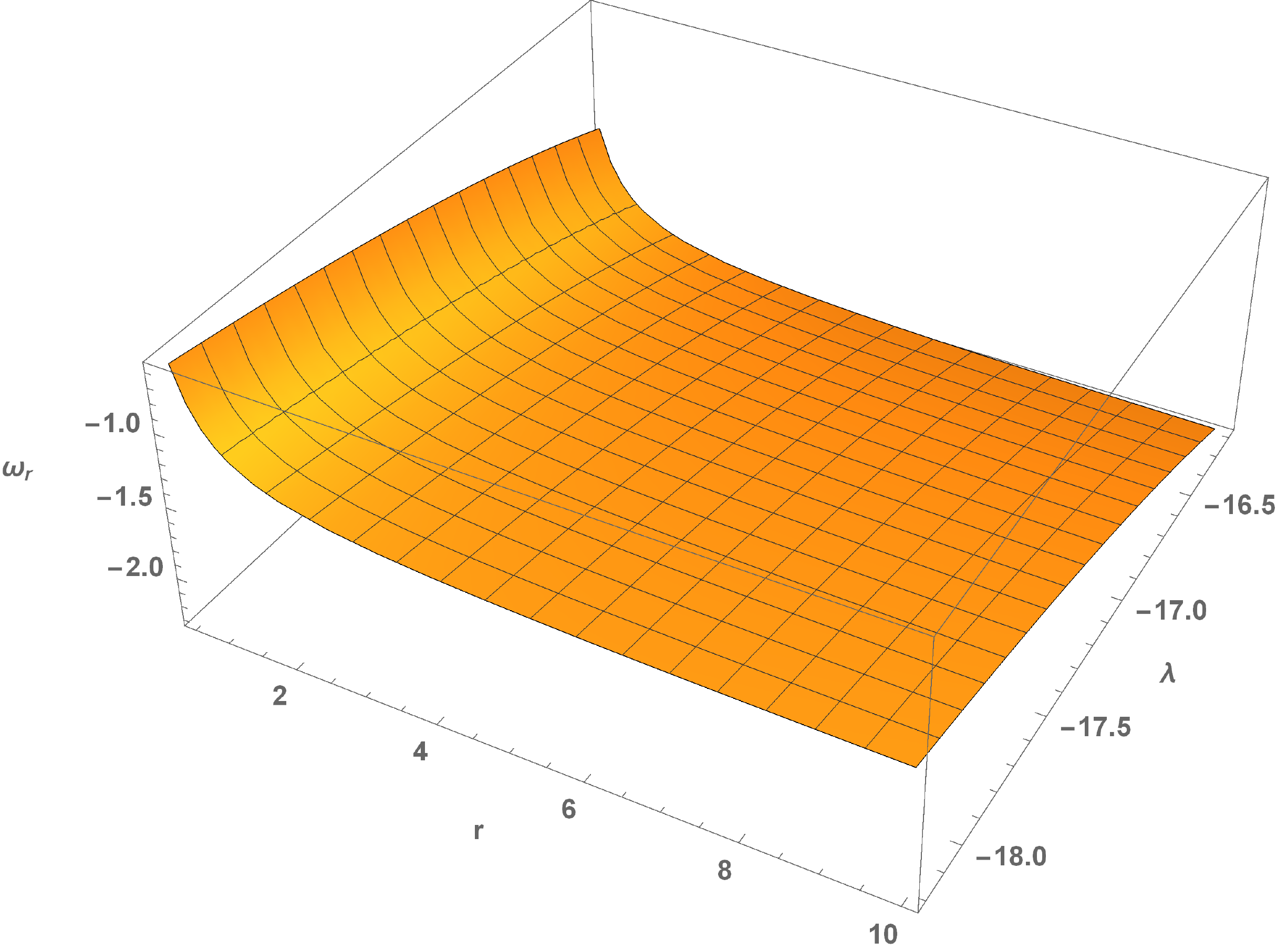}
\caption{Radial equation of state parameter $\omega_r=p_r/\rho$ as a function of $r$ and $\lambda$ with $r_0 = 0.5$.}\label{fig10}
\end{figure}

\begin{figure}[H]
\includegraphics[scale =0.28]{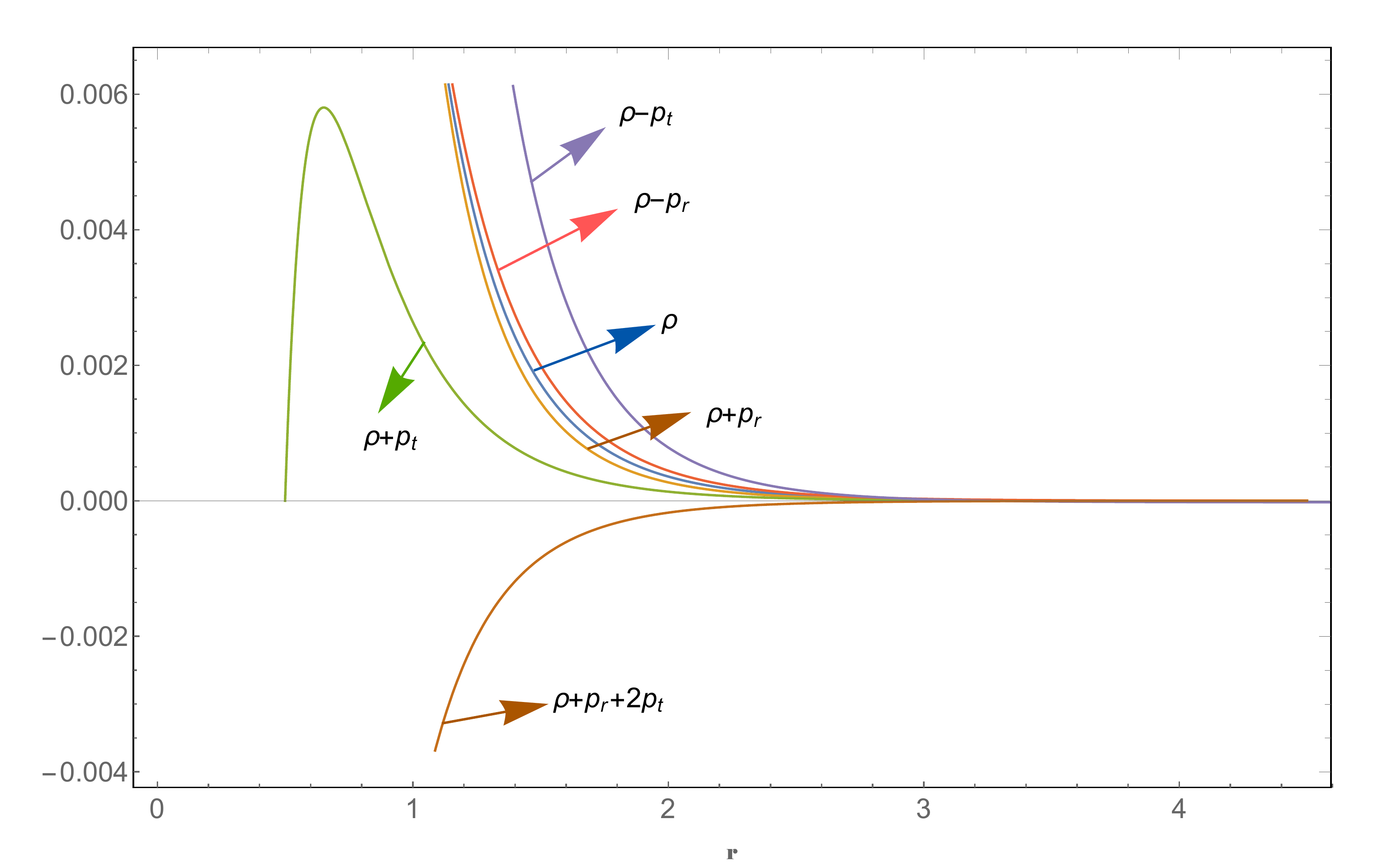}
\caption{The energy conditions as a function of $r$ with $\lambda=-30$ and $r_0 = 0.5$.}\label{fig11}
\end{figure}

\begin{figure}[H]
\includegraphics[scale =0.28]{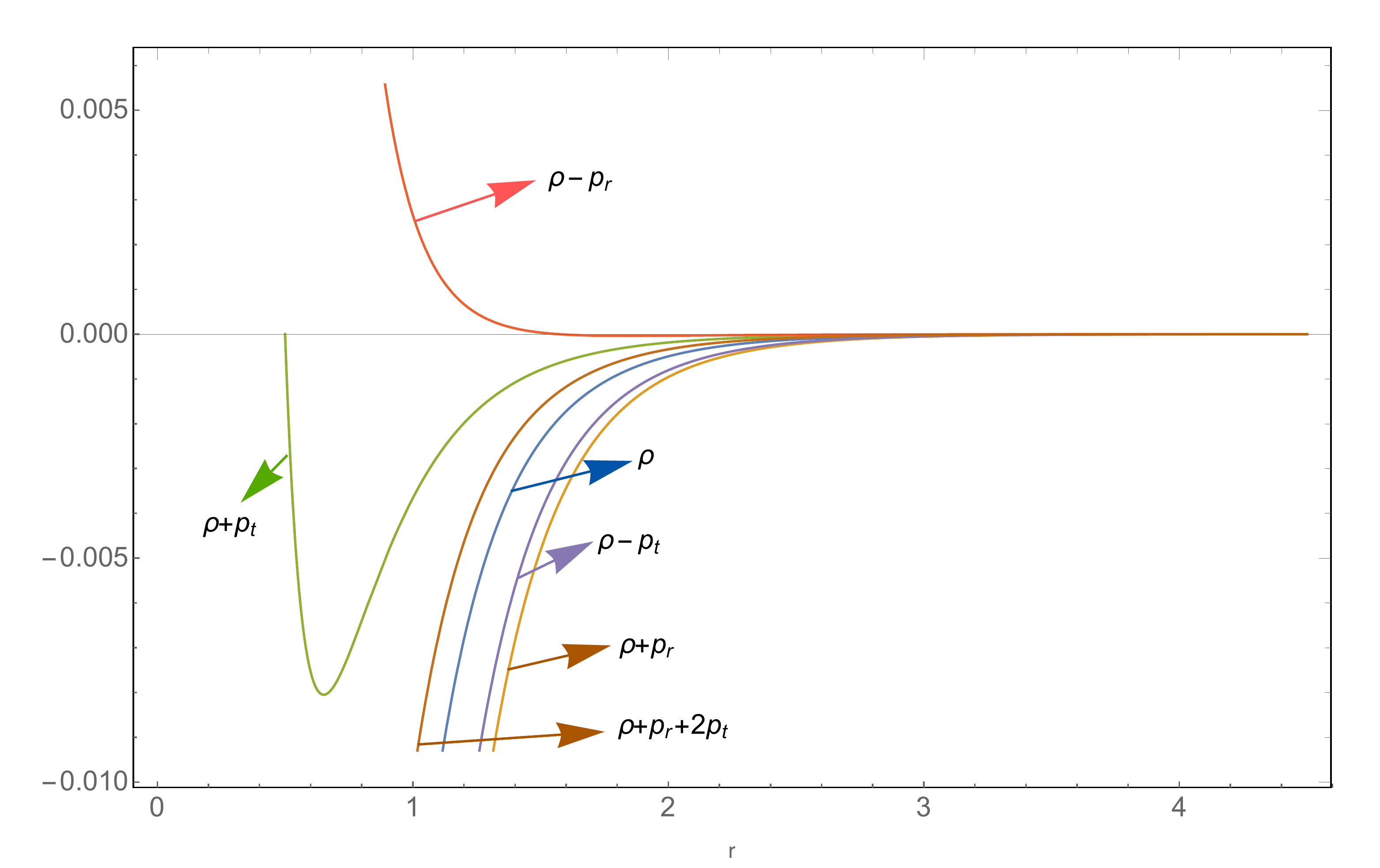}
\caption{The energy conditions as a function of $r$ with $\lambda=0$ and $r_0 = 0.5$.}\label{fig12}
\end{figure}

Here, it is determined that  $\lambda$  must remain in the range of - 80 to - 13 to validate NEC, WEC and DEC.  In this article we have considered $\lambda=-30$. The respective behaviour in GR is plotted in Fig.\ref{fig12}.

It is quite clear that in the absence of the $f(R,T)$ gravity extra terms, most of the energy conditions are no longer respected. In fact, only one of the energy conditions is respected even within the assumption of such a promising and fruitful proposal for the WH shape function.

The $f(R,T)$ gravity, departing from many alternative gravity theories in the present literature, allows one to modify the effective energy-momentum tensor worked out in GR. This is made pretty clear in Eq.(5), in which even if one assumes the energy-momentum tensor of a perfect fluid, the effective energy-momentum tensor of the theory, namely,

\begin{equation}
T_{ij}^\text{eff}=T_{ij}+\frac{\lambda}{8\pi}[Tg_{ij}+2(T_{ij}+pg_{ig})],
\end{equation}
presents ``imperfect'' fluid terms, that may be related to viscosity or anisotropy. Such a strong modification in the way one sees the effective fluid permeating a particular astrophysical or cosmological system, such as in the present case the wormhole, makes possible to obtain significantly different material features and in our case, consequently, makes possible to respect the energy conditions, as a consequence of the description of viscosity/anisotropy disguised in terms proportional to $\lambda$.

\section{Conclusion}\label{VII}

WHs are tube-like structures which, as shortcuts, connect two distant regions in the universe (or even in different universes). If their geometrical structure was not singular enough, according to GR formalism, WHs are expected to be filled by exotic (negative mass) matter. 

The lack of observations of WHs so far makes one unable to predict exactly some of their geometrical and material properties such as the shape function and the EoS. In the present article we have proposed a novel functional form for the shape function, which depends only exponentially on $r$. On the other hand we did not need to assume any particular form for the WH EoS, which was obtained from the model, rather than imposed to it, as it happens in some cases in the literature \cite{lobo/2005}-\cite{ms/2017}.

Before going any further on the discussion of the WH EoS obtained, we should mention that for the exponential shape function presented in Eq.\eqref{e14}, Fig.\ref{fig9} has shown that it satisfies all the requirements needed to have traversable asymptotically flat WHs. In this way we were allowed to obtain the EoS parameter solution $\omega_r$ as well as to construct the WH energy conditions.

Fig.\ref{fig10} shows that the WH EoS is in the phantom region, that is, $\omega_r<-1$. It is well known that a phantom  EoS parameter $<-1$ for the universe will imply in the so-called Big Rip \cite{caldwell/2003} though some alternative to evade such a catastrophic scenario have appeared \cite{wu/2005}-\cite{msts/2019}. Phantom WHs have also appeared in the literature \cite{lobo/2005}-\cite{lobo/2005b} though it is important to remark that in these cases the phantom EoS was invoked rather than obtained from the model, as in the present case.

Fig.\ref{fig11} show the energy conditions of the present WHs scenario. They show a properly obedience of NEC and WEC, departing from standard GR solutions. The DEC is also satisfied while SEC is not. 

Similar approaches to WHs in $f(R,T)$ gravity can be seen in the literature, though with non-exponential shape functions \cite{elizalde/2019,ms/2017}, \cite{zubair/2019}-\cite{azizi/2013}. By comparing our approach with those, the important role of the exponential shape function becomes clear since none of them presents WHs fully satisfying NEC, WEC and DEC, as our model does (recall Fig.1). 

The SEC, which is not obeyed in the present article as well as in many others, is a subject under discussion for some time. For instance, the SEC must be violated during the inflationary epoch and the need for this violation is why inflationary models are typically driven by scalar inflation fields \cite{Visser/2000}. Further, the recent observational data regarding the accelerating universe \cite{riess/1998}-\cite{tonry/2003} makes the SEC to be violated on cosmological scales right now \cite{Visser/1997}.

\acknowledgments{PHRSM would like to thank S\~ao Paulo Research Foundation (FAPESP), grant 2018/20689-7, for financial support. PKS acknowledges DST, New Delhi, India for providing facilities through DST-FIST lab, Department of Mathematics, where a part of this work was done. We are very much grateful to the honorable referees and the editor for the illuminating suggestions that have significantly improved our work in terms of research quality and presentation.}

\end{document}